\documentstyle[aps,prl,floats,psfig,epsf,graphicx]{revtex}

\begin{document}
\draft

\wideabs{

\title{Mott-Hubbard Metal-Insulator Transition in Paramagnetic
$ \rm {\bf V_2O_3} $:\\  a LDA+DMFT(QMC) Study}

\author{K.\ Held$^{1,*}$, G.\ Keller$^1$, V.\ Eyert$^1$, D.\ Vollhardt$^1$, and V.\ I.\ Anisimov$^2$}
\address{$^1$ Institut f\"ur Physik, Universit\"at Augsburg,
         86135 Augsburg, Germany}
\address{$^2$ Institute of Metal Physics, Ekaterinburg GSP-170, Russian}
 \date{\today}

\maketitle

\begin{abstract}
\noindent The electronic properties of paramagnetic $\rm V_2O_3$
are investigated by the {\em ab-initio} computational scheme
LDA+DMFT(QMC). This approach merges the local density
approximation (LDA) with dynamical mean-field theory (DMFT) and
uses numerically exact quantum Monte Carlo simulations (QMC) to
solve the effective Anderson impurity model of DMFT. Starting with
the crystal structure of metallic  $ {\rm V_2O_3} $ and insulating
$ {\rm (V_{0.962}Cr_{0.038})_2O_3}$ we find a Mott-Hubbard
metal-insulator-like transition
 at a Coulomb interaction $U\approx5$ eV. The calculated spectrum is in very good agreement with
experiment. Furthermore, the occupation of the ($a_{1g}$,
$e_{g1}^{\pi }$, $e_{g2}^{\pi }$) orbitals and the spin state
$S=1$ determined by us agree with recent polarization dependent
X-ray-absorption experiments. 
\end{abstract}
}

The metal-insulator transition within the paramagnetic phase of ${\rm %
V_{2}O_{3}}$ is generally considered to be the classical example
of a Mott-Hubbard metal-insulator transition (MIT)\cite{Mott}.
During the last few years, our understanding of the MIT in the
one-band Hubbard model has considerably improved, in particular
due to the application of dynamical mean-field theory (DMFT).

Within DMFT the electronic lattice problem is mapped onto a
self-consistent single-impurity Anderson model. The mapping
becomes exact in the limit of infinite coordination
number\cite{DMFT} and allows one to investigate the dynamics of
correlated lattice electrons non-perturbatively
at all interaction strengths. This is of essential
importance for a problem like the MIT which occurs at a Coulomb
interaction comparable to the electronic band-width. The
transition is signaled by the collapse of the quasiparticle peak
at the Fermi energy when the interaction is
increased\cite{DMFTMott}. Rozenberg {\it et al.} \cite{Rozenberg95}
first applied DMFT to investigate the metal-insulator transition
in ${\rm V_{2}O_{3}}$ in terms of the one-band Hubbard model.
Subsequently, the influence of orbital degeneracy was studied by
means of the two-\cite{Rozenberg97a,Han98a,Held98a} and
three-band\cite{Han98a} Hubbard model for a Bethe density of
states (DOS).

Clearly, a realistic theory of ${\rm V_{2}O_{3}}$ must take into
account the complicated electronic structure of this system. In
the high-temperature paramagnetic phase ${\rm V_{2}O_{3}}$ has a
corundum crystal structure in which the $V$ ions are surrounded by
an octahedron of oxygen ions. This leads to an electronic
structure with a $3d^{2}$ $V^{3+}$ state, where the two
$e_{g}$-orbitals are empty and \ the three $t_{2g}$-orbitals are
filled with two electrons. A small trigonal distortion lifts the
triple degeneracy of the $t_{2g}$-orbitals, resulting in one
non-degenerate $a_{1g}$-orbital oriented along the c-axis and two
degenerate $e_{g}^{\pi }$ orbitals oriented predominantly in the
hexagonal plane. Starting from this orbital structure Castellani
{\it et al.} \cite{castelani} proposed a widely accepted model with a
strong covalent $a_{1g}$-bond between two V ions along the c-axis.
This bonding state is occupied by a singlet pair (one electron per
V) and hence does not contribute to the local magnetic moment. The
remaining electron per $V$ has a twofold orbital degeneracy within
the $e_{g}^{\pi }$ orbitals and a spin $S=1/2$. This $S=1/2$ model
suggested that the half filled, one-band Hubbard model was the
simplest possible electronic model describing ${\rm V_{2}O_{3}}$.
However, recent experimental results by Park {\em et
al.}\cite{park} obtained by polarized X-ray spectroscopy seem to
require an interpretation in terms of a $S=1$ spin state, and an
$e_{g}^{\pi }e_{g}^{\pi }$ orbital state with an admixture of
$e_{g}^{\pi }a_{1g}$ configurations. Subsequently, Ezhov {\em et
al.} \cite{ezhov99} and Mila {\em et al.} \cite{Mila00} argued for
$S=1$ models without and with orbital degeneracy, respectively,
for the antiferromagnetic insulating phase of ${\rm V_{2}O_{3}}$.
LDA+U calculations indicate that the atomic Hund's rule is
responsible for the high-spin ground state of the V
ions\cite{ezhov99}. While LDA+U may be used to describe the
antiferromagnetic insulating phase of ${\rm V_{2}O_{3}}$, the
metal-insulator transition within the correlated {\em
paramagnetic} phase is beyond the scope of this approach since the
Coulomb interaction is treated within Hartree-Fock. Here the
computational scheme LDA+DMFT
\cite{Anisimov97a,Lichtenstein98a,Zoelfl00,Nekrasov00}, obtained
by combining electronic bandstructure theory (LDA) with the
many-body technique DMFT \cite{DMFT}, is the best available {\em
ab initio} method for the investigation of correlated electronic
systems close to a Mott-Hubbard MIT\cite{Held00}. To solve the
DMFT-equations we employ quantum Monte-Carlo simulations (QMC)
which yield a numerically exact solution\cite{Foot2}; the
resulting calculational scheme is referred to as
LDA+DMFT(QMC)\cite{Lichtenstein98a,Nekrasov00,Held00}.

\begin{figure}[htb]
\centerline{
\includegraphics[clip=true,width=8.5cm]{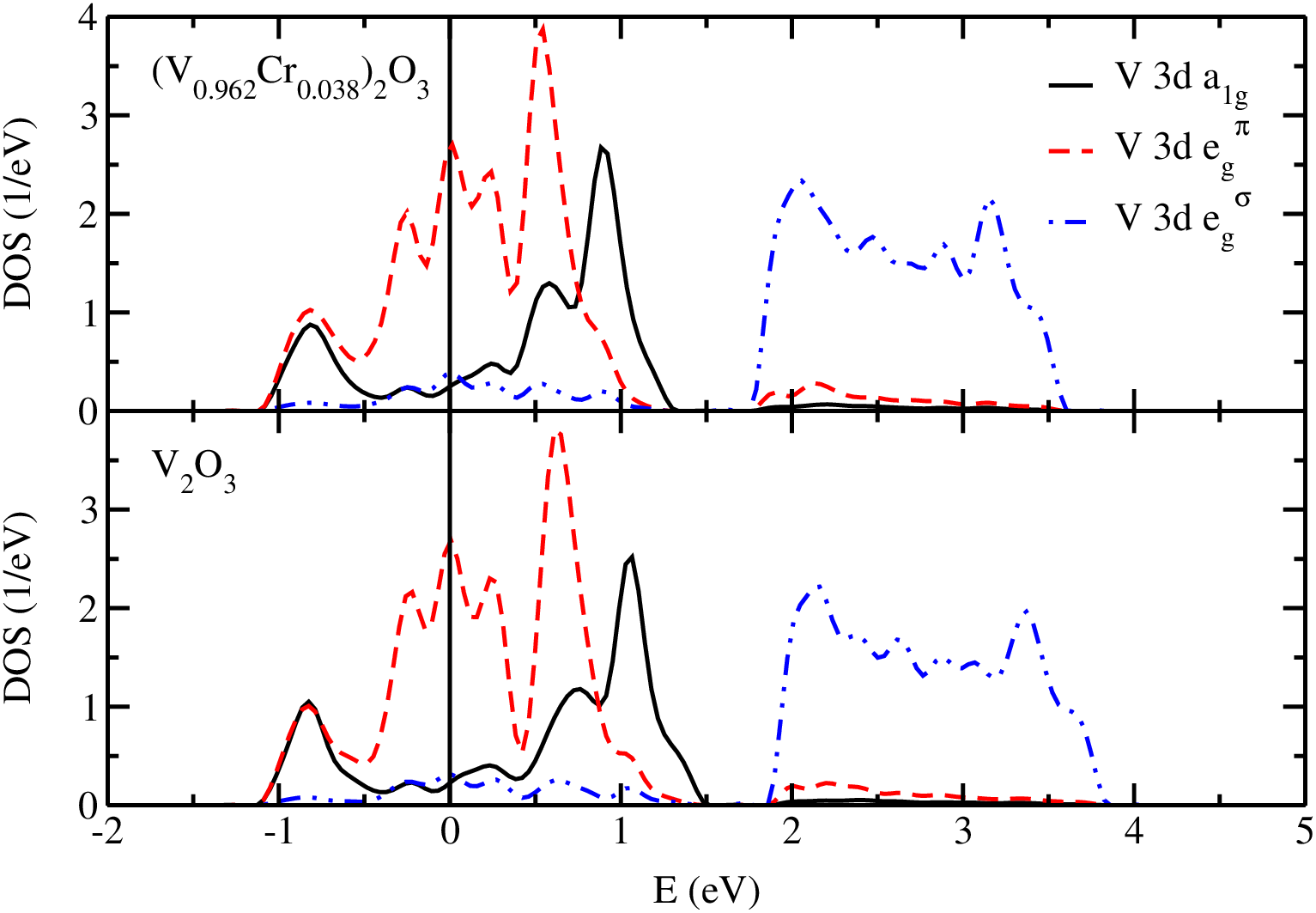} }
\vspace{1em} \caption{Partial DOS of the 3d bands  per unit cell
for paramagnetic a) metallic $ {\rm V_2O_3}$ and b) insulating $
{\rm (V_{0.962}Cr_{0.038})_2O_3}$.} \label{ldados}
\end{figure}

We start by calculating the density of states (DOS) of
paramagnetic
{\em metallic} ${\rm V_{2}O_{3}}$ and paramagnetic {\em insulating} ${\rm %
(V_{0.962}Cr_{0.038})_{2}O_{3}}$, respectively, within LDA using
the crystal structure data of Dernier\cite{dernier70a,Foot1}.
In the LDA DOS, Fig.\ \ref{ldados},
one observes the expected behavior: the V $t_{2g}$ states are near
the Fermi energy and are split into a non-degenerate $a_{1g}$ band
and doubly degenerate $e_{g}^{\pi }$ bands; the\ V $e_{g}^{\sigma
}$ states are at
higher energies. The results for corundum ${\rm V_{2}O_{3}}$ and ${\rm %
(V_{0.962}Cr_{0.038})_{2}O_{3}}$ are very similar, and are close
to those
found by Mattheiss\cite{mattheiss94} and Ezhov {\em et al.}\cite%
{ezhov99}\ for the corundum and the monoclinic phase. The change
in the LDA
DOS on going from metallic ${\rm V_{2}O_{3}}$ to insulating $({\rm %
V_{0.962}Cr_{0.038})_{2}O_{3}}$ only consists in a slight narrowing of the $%
t_{2g}$ and $e_{g}^{\sigma }$ bands by $\approx 0.2$ and $0.1$ eV,
respectively, as well as a weak downshift of the centers of
gravity of both groups of bands, which can be attributed to the
crystallographic changes. Most important is the fact that the
insulating gap observed experimentally for the Cr-doped system is
{\em missing} in the LDA DOS. It is generally believed that this
insulating gap is due to strong Coulomb interactions which cannot
be described adequately within LDA. Using LDA+DMFT(QMC) we will
now show explicitly that the insulating gap is indeed caused by
the electronic correlations.

In the LDA+DMFT
approach\cite{Anisimov97a,Lichtenstein98a,Zoelfl00,Nekrasov00,Held00}
the LDA band structure, expressed by a one-particle Hamiltonian
$H_{{\rm LDA}}^{0}$, is supplemented with the local Coulomb
repulsion $U$ and Hund's rule exchange $J$:
\begin{eqnarray}
\hat{H} &=&\hat{H}_{{\rm LDA}}^{0}+{\ U}\sum_{m}\sum_{i}\hat{n}_{im\uparrow }%
\hat{n}_{im\downarrow }  \nonumber \\ &&+\;\sum_{i\;m\neq
\tilde{m}\;\sigma \tilde{\sigma}}\;(V-\delta _{\sigma
\tilde{\sigma}}J)\;\hat{n}_{im\sigma
}\hat{n}_{i\tilde{m}\tilde{\sigma}}. \label{H}
\end{eqnarray}%
Here, $i$ denotes the lattice site; $m$ and $\tilde{m}$ enumerate
the three
interacting $t_{2g}$ orbitals. The interaction parameters are related by $%
V=U-2J$ which holds exactly for degenerate orbitals and is a good
approximation in our case. Furthermore, since the $t_{2g}$ bands
at the Fermi energy are rather well separated from all other bands
we restrict the calculation to these bands (for
details of the computational scheme see Refs.\cite{Nekrasov00,Held00}). Consequently, only the LDA DOS of the three $%
t_{2g}$ bands shown in Fig.\ref{ldados} enter the
calculation\cite{Held00}. While the Hund's rule coupling $J$ is
insensitive to screening effects and
may thus be obtained within LDA to a good accuracy ($J=0.93$ eV \cite%
{Solovyev}), the LDA-calculated value of the Coulomb repulsion $U$
has a typical uncertainty of at least 0.5 eV\cite{Nekrasov00}.
Hence, the critical value of $U$ for the MIT in ${\rm V_{2}O_{3}}$
is not {\em a priori} known and must be determined within our
scheme.

\begin{figure}[htb]
\centerline{\hbox{
\includegraphics[clip=true,width=8.5cm]{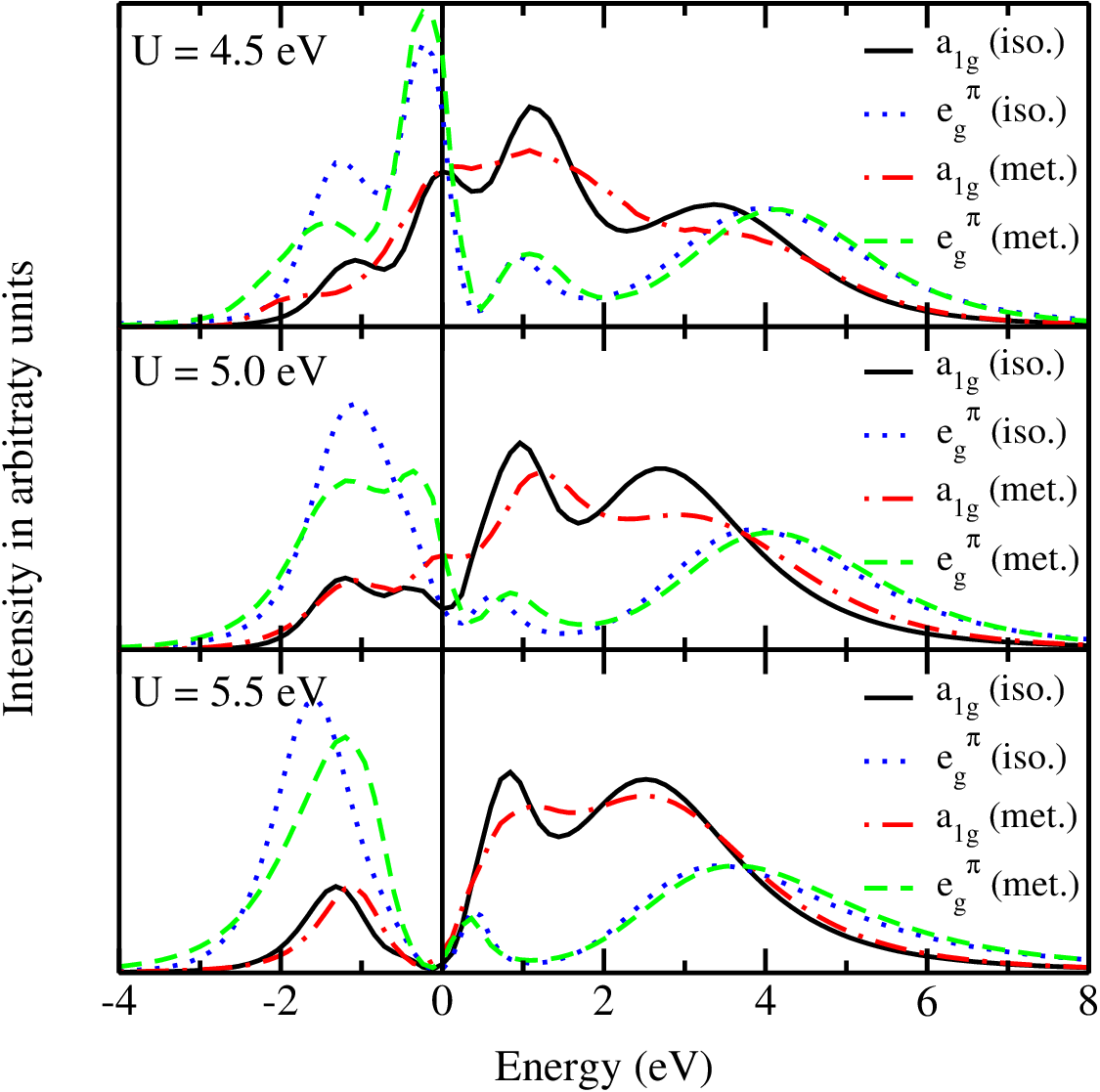} }}
\vspace{1em} \caption{LDA+DMFT(QMC) spectra for paramagnetic  $
{\rm (V_{0.962}Cr_{0.038})_2O_3}$ (``iso.'') and ${\rm V_2O_3} $
(``met.'') at $U=4.5$, $5$ and $5.5$ eV, and $T=0.1$ eV $\approx
1000$ K.} \label{spectrum}
\end{figure}

The spectra obtained by LDA+DMFT(QMC) are shown in
Fig.\ref{spectrum}\cite{Foot3}. They imply that the critical value
of $U$ for the MIT is about $5$ eV. Indeed, at $U=4.5$ eV one
observes pronounced quasiparticle peaks at the Fermi energy, i.e.,
characteristic metallic behavior, even for the crystal structure of ${\rm %
(V_{0.962}Cr_{0.038})_{2}O_{3}}$, while at $U=5.5$ eV the form of
the calculated spectral function is typical for an insulator for
both sets of crystal structure parameters. At $U=5.0$ eV one is
then at, or very close to, the MIT since there is a pronounced dip
in the DOS at the Fermi energy
for both $a_{1g}$ and $e_{g}^{\pi }$ orbitals for the crystal structure of $%
{\rm (V_{0.962}Cr_{0.038})_{2}O_{3}}$, while for pure ${\rm
V_{2}O_{3}}$ one still finds quasiparticle peaks.
(We note that at $1000$K one only observes metallic-like and
insulator-like behavior, with a rapid but smooth\ crossover
between these two phases, since a sharp MIT occurs only at lower
temperatures\cite{Rozenberg97a,DMFTMott}).
The critical value of the Coulomb interaction $U\approx 5$ eV is
in reasonable agreement with the values determined
spectroscopically by fitting to model calculations, and by
constrained LDA. The former gives $U=4-5$ eV for vanadium oxides
\cite{U-exp} while the latter
yields \cite{Solovyev} $U\approx 3$ eV (for ${\rm LaVO_{3}}$) to $U\approx 8$%
eV, depending on whether the $e_{g}$-electrons participate in the
screening
or not; without screening, one finds $U\approx 6-7$ eV \cite%
{Anisimov91}.

\begin{figure}[htb]
\centerline{\hbox{
\includegraphics[clip=true,width=8.5cm]{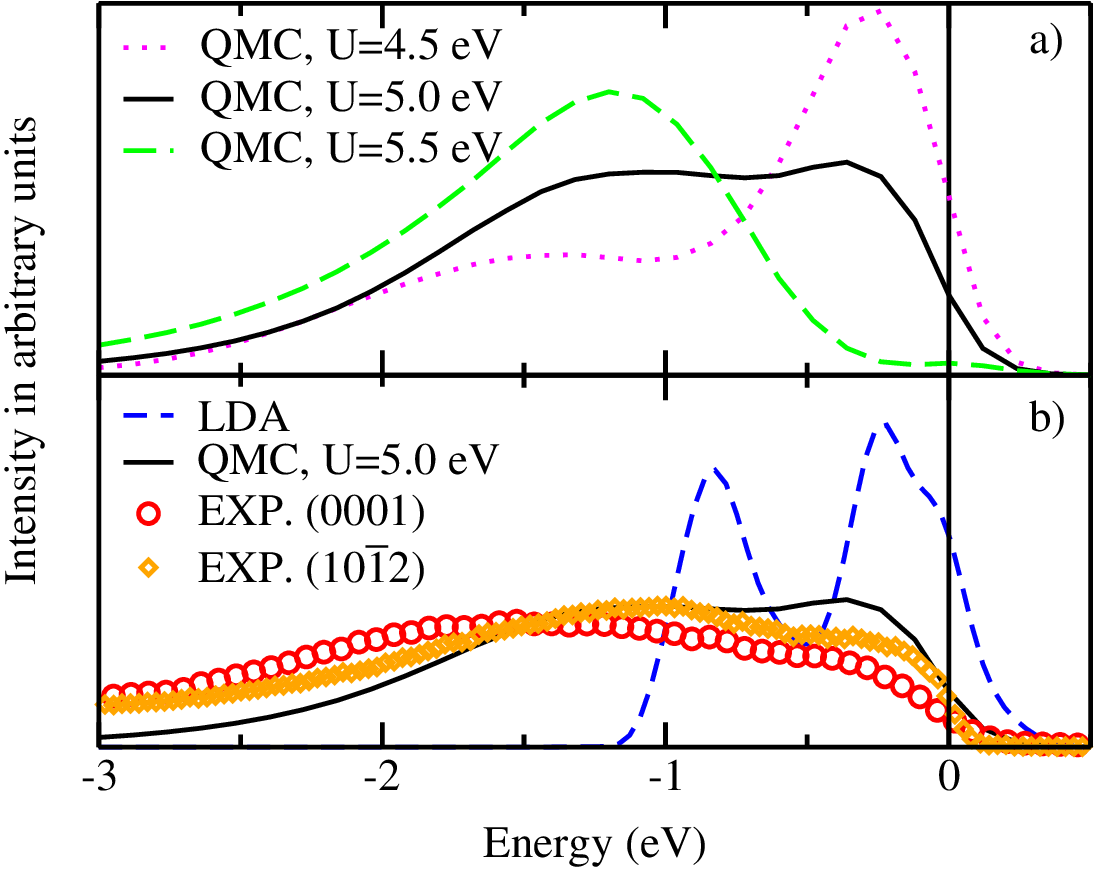} }}
\vspace{1em} \caption{a) LDA+DMFT(QMC) spectrum for $U=4.5$, $5$
and $5.5$ eV at $T=1000$ K;  b) Comparison  with the LDA spectrum
and the photoemission experiment by Schramme {\it et al.} \protect
\cite{Schramme00} for two different ${\rm V_2O_3} $ single-crystal
surfaces at $T=300$ K. Note, that the (10$\bar{1}$2) surface has 
the same coordination number as the bulk. } \label{compexp}
\end{figure}

To compare with the photoemission spectrum of ${\rm V_{2}O_{3}}$
by Schramme
{\em et al.}\cite{Schramme00}, the LDA+DMFT(QMC) spectra of Fig.\ref%
{spectrum} are multiplied with the Fermi function at $1000$ K and
Gauss-broadened by $0.05$ eV to account for the experimental
resolution. The theoretical results for $U=5$ eV are seen to be in
good agreement with experiment (Fig.~\ref{compexp}), in contrast
with the LDA results.
We also note that the DOS is highly asymmetric w.r.t the Fermi
energy due to the orbital degrees of freedom. The comparison
between our results, the data of M\"{u}ller {\it et al.}
\cite{Mueller} obtained by X-ray absorption measurements, and LDA
in Fig. 4 shows that, in contrast with LDA, our results not only
describe the different bandwidths above and below the Fermi energy
($\approx 6$ eV and $\approx 2-3$ eV, respectively) correctly, but
even resolve the two-peak structure above the Fermi energy.
Also the interpretation of the two peaks is different in LDA and
LDA+DMFT(QMC). While in the latter approach the peak at lower
energies ($1$ eV) has predominantly $a_{1g}$ character and the
peak at higher energy ($3$ eV) is due to the $e_{g}^{\pi }$
Hubbard band (see Fig. 2), in LDA the $a_{1g}$ and $e_{g}^{\pi }$
states contribute only below $1.8$ eV such that the second peak is
entirely due to the $e_{g}^{\sigma}$ states (see Fig. 1).
\begin{figure}[htb]
\centerline{\hbox{
\includegraphics[clip=true,width=8.5cm]{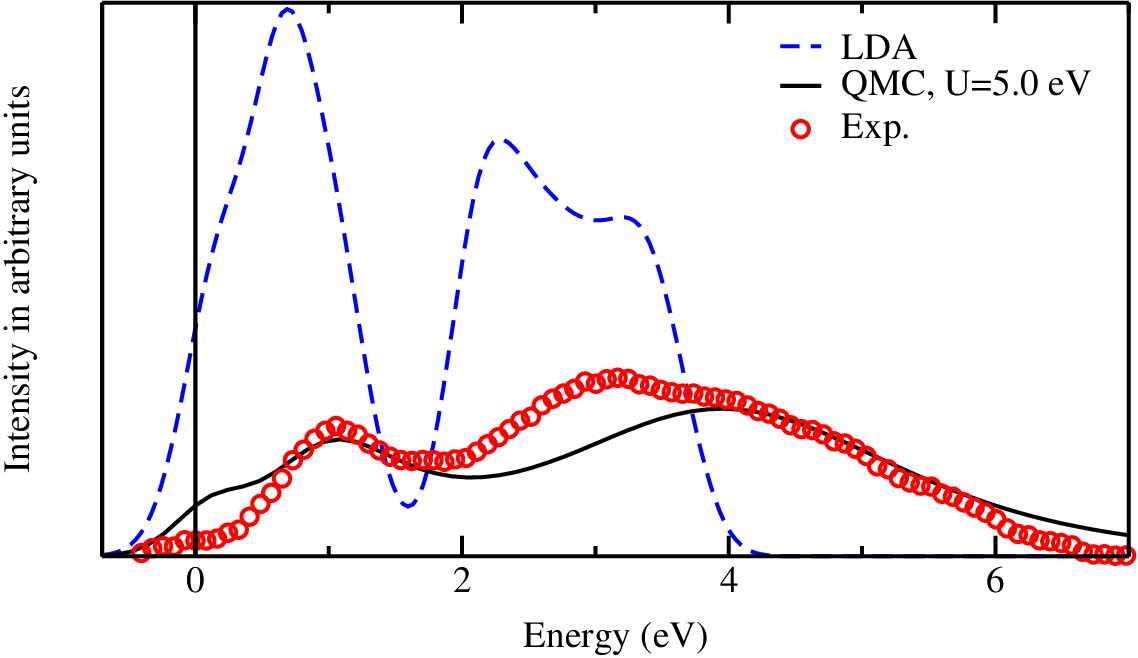} }}
\vspace{1em} \caption{Comparison of the LDA and LDA+DMFT(QMC)
spectra at $T=1000$K (Gaussian broadened with $0.2$ eV) with the X
-ray absorption data of M\"uller {\it et al.}  \protect
\cite{Mueller}. The LDA and QMC curves are normalized differently
since the $e_{g}^{\sigma}$ states, which are shifted towards
higher energies when the Coulomb interaction is included, are
neglected in the QMC calculation.
 } \label{spectrum2}
\end{figure}
Particularly interesting are the spin and the orbital degrees of
freedom in ${\rm V_{2}O_{3}}$. We find (not shown) that for
$U\gtrsim 3$ eV the squared local magnetic moment $\left<m_{z}^{2}\right>=
\left<\left(\sum_{m} 
[\hat{n}_{m\uparrow}-\hat{n}_{m\downarrow}]\right)^2\right>$
saturates at a value of $4$, i.e., there are {\em two} electrons
with the same spin direction in the ($a_{1g}$, $e_{g1}^{\pi }$,
$e_{g2}^{\pi }$) orbitals. Thus we conclude that the spin state of
${\rm V_{2}O_{3}}$ is $S=1$ throughout the Mott-Hubbard transition
region. This agrees with the
measurements of Park {\em et al.}%
\cite{park} and also with the data for the high-temperature
susceptibility \cite{S-exp}. The latter correspond to an effective
magnetic moment $\mu _{eff}=2.66\mu _{B}$ which is close to the
ideal value $\mu _{eff}=2.83\mu _{B}$ for $S=1$. The result is at
odds with a $S=1/2$ model and with the results for a one-band
Hubbard model where $m_{z}^{2}$ changes substantially at the MIT
\cite{DMFTMott}.

For the orbital degrees of freedom we find a predominant occupation of the $%
e_{g}^{\pi }$ orbitals, but with a significant admixture of
$a_{1g}$ orbitals. This admixture decreases at the MIT: in the
metallic phase we determine the occupation of the ($a_{1g}$,
$e_{g1}^{\pi }$, $e_{g2}^{\pi }$) orbitals as (0.37, 0.815,
0.815), and in the insulating phase as (0.28, 0.86, 0.86). This
should be compared with the experimental results of Park {\it et
al.} \cite{park}. From their analysis of the linear dichroism data
the authors concluded that the ratio of the configurations
$e_{g}^{\pi }e_{g}^{\pi }$:$e_{g}^{\pi }a_{1g}$ is equal to 1:1
for the paramagnetic metallic and 3:2 for the paramagnetic
insulating phase, corresponding to a one-electron occupation of
(0.5,0.75,0.75) and (0.4,0.8,0.8), respectively.
Although our results show a somewhat smaller value for the admixture of $%
a_{1g}$ orbitals, the overall behavior, including the tendency of a {\em %
decrease }of the $a_{1g}$ admixture across the transition to the
insulating
state, are well reproduced. This
agrees with the LDA+U calculation for the
antiferromagnetic insulating phase\cite{ezhov99}. Further
experimental evidence for predominant $(e_{g}^{\pi },e_{g}^{\pi
})$ configurations was obtained by polarized neutron diffraction
by Brown {\it et al.}\cite{brown}. Measuring the spatial
distribution of the magnetization induced by a magnetic field,
they showed that the moment induced on the V atoms is almost
entirely due to the electrons in the doubly degenerate $e_{g}^{\pi
}$ orbitals.

In conclusion, starting from the LDA-calculated spectra for paramagnetic $%
{\rm V_{2}O_{3}}$ and ${\rm (V_{0.962}Cr_{0.038})_{2}O_{3},}$ and
including the missing electronic correlations via DMFT(QMC), we
showed that a Mott-Hubbard MIT occurs in the paramagnetic phase at
$U\approx 5$eV. Our results are in very good agreement with the
experimentally determined photoemission and X-ray absorption
spectra, i.e., above {\em and} below the Fermi energy.
Furthermore, we calculate the spin state to be $S=1$ and find an
orbital admixture of $e_{g}^{\pi }e_{g}^{\pi }$ and $e_{g}^{\pi
}a_{1g}$ configurations, which both agree with recent experiments.
Thus LDA+DMFT(QMC) provides a remarkably accurate microscopic
theory of the strongly correlated electrons in the paramagnetic
phase of ${\rm V_{2}O_{3}}$.

The MIT will eventually become first order at lower temperatures
\cite{Rozenberg95,DMFTMott}; QMC simulations at $T\approx 300$ K
are under way, but are very computer-expensive. Furthermore,
future {\em ab initio} investigations will have to clarify the
origin of the discontinuous lattice distortion at the first-order
MIT which leaves the lattice symmetry unchanged. Here various
scenarios are possible\cite{Majumdar94,KVC}. In particular, the
MIT might be the driving force behind the lattice distortion by
causing a thermodynamic instability w.r.t. changes of the lattice
volume and distortions.


We acknowledge valuable discussions with N. Bl\"{u}mer, R.
Claessen, U. Eckern, K.-H. H\"{o}ck, S. Horn, and S.A. Sawatzky. This work was supported by
the Deutsche Forschungsgemeinschaft through SFB 484 and
Forschergruppe HO 955/2, by the Russian Foundation for Basic
Reasearch grant RFFI-98-02-17275, by the John von Neumann-Institut
f\"{u}r Computing, J\"{u}lich, and by a Feodor-Lynen
grant of the Alexander von Humboldt-Foundation (KH).  %
\vspace{-1.5em}
%
%
%


\begin{references}
\vspace{-2em}

\bibitem[*]{} Present address: Physics Department, Pinceton Univer-\newline
sity,~Princeton,~NJ 08544;~kheld@feynman.princeton.edu.

\bibitem{Mott} N. F. Mott, Rev.\ Mod.\ Phys.\ {\bf 40}, 677 (1968); {\sl %
Metal-Insulator Transitions} (Taylor \& Francis, London, 1990); F.
Gebhard, {\sl The Mott Metal-Insulator Transition} (Springer,
Berlin, 1997).

\bibitem{DMFT} W. Metzner and D. Vollhardt, Phys.\ Rev.\ Lett.\ {\bf 62}, 324
(1989); A. Georges, G. Kotliar, W. Krauth, and M. J. Rozenberg,
Rev.\ Mod.\ Phys.\ {\bf 68}, 13 (1996).

\bibitem{DMFTMott} G. Moeller, Q. Si, G. Kotliar, M. Rozenberg, and
D. S. Fisher, Phys.\ Rev.\ Lett.\ {\bf 74},
2082 (1995); J. Schlipf, M. Jarrell, P. G. J. van Dongen, N. Bl{\"{u}}mer,
 S. Kehrein, Th. Pruschke, and D. Vollhardt, Phys.\ Rev.\ Lett.\ {\bf
82}, 4890 (1999); M. J. Rozenberg, R. Chitra and G. Kotliar,
Phys.\ Rev.\ Lett.\ {\bf 83}, 3498 (1999); R. Bulla, Phys.\ Rev.\
Lett.\ {\bf 83}, 136 (1999).

\bibitem{Rozenberg95} M. J. Rozenberg, G. Kotliar, H. Kajueter, G. A. Thomas,
 D. H. Rapkine, J. M. Honig, and P. Metcalf, Phys.\ Rev.\ Lett.\ {\bf 75}%
,105 (1995)

\bibitem{Rozenberg97a} M.~J.~Rozenberg, Phys.\ Rev.\ B {\bf 55}, R4855 (1997).

\bibitem{Han98a} J. E. Han, M. Jarrell, and D. L. Cox, Phys. Rev. B {\bf 58}%
, R4199 (1998).

\bibitem{Held98a} K. Held and D. Vollhardt,
\newblock { Euro. Phys. J. B \bf
5}, 473 (1998).

\bibitem{castelani} C. Castellani, C. R. Natoli, and J. Ranninger,
 Phys.\ Rev.\ B {\bf 18}, 4945
(1978); {\bf 18}, 4967 (1978); {\bf 18}, 5001 (1978).

\bibitem{park} J.-H. Park, L.H. Tjeng, A. Tanaka, J.W. Allen, C.T. Chen,
P. Metcalf, J.M. Honig, F.M.F. de Groot, and S.A. Sawatzky,
Phys.\ Rev.\ B {\bf 61}, 11 506 (2000).

\bibitem{ezhov99} S.\ Yu.\ Ezhov,
V.\ I.\ Anisimov, D.\ I.\ Khomskii, and G.\ A.\ Sawatzky,
Phys.\ Rev.\ Lett.\ {\bf 83}, 4136 (1999).

\bibitem{Mila00} F. Mila, R. Shiina, F.-C. Zhang, A. Joshi, M. Ma,
V. Anisimov, and T. M. Rice, Phys.\ Rev.\ Lett.\ {\bf 85}, 1714
(2000).

\bibitem{Anisimov97a} V. I. Anisimov,
A. I. Poteryaev, M. A. Korotin, A. O. Anokhin, and G. Kotliar, J.
Phys.: Cond. Matt. {\bf 9}, 7359 (1997).

\bibitem{Lichtenstein98a} A. I. Lichtenstein and M. I. Katsnelson, Phys.
Rev. B {\bf 57}, 6884 (1998).

\bibitem{Zoelfl00}  M. B. Z\"{o}lfl, Th. Pruschke, J. Keller, A. I. Poteryaev,
I. A. Nekrasov, and V. I. Anisimov, Phys. Rev. B {\bf 61}, 12810
(2000).

\bibitem{Nekrasov00} I. A. Nekrasov, K. Held, N. Bl{\"{u}}mer,
 A. I. Poteryaev,  V. I. Anisimov, and D. Vollhardt,
Euro Phys. J. B {\bf 18}, 55 (2000).

\bibitem{Held00} For an introduction into LDA+DMFT, see K. Held, I. A. Nekrasov, N. Bl{\"{u}}mer,
V. I. Anisimov, and D. Vollhardt,
preprint cond-mat/00010395.

\bibitem{Foot2} For a comparison of LDA+DMFT results for the photoemission
spectra of ${\rm La_{1-x}Sr_{x}TiO_{3}}$ obtained by different
numerical techniques see Ref.\cite{Nekrasov00}.

\bibitem{dernier70a} P.\ D.\ Dernier, J.\ Phys.\ Chem.\ Solids {\bf 31},
2569 (1970).

\bibitem{Foot1} Use of the crystal structure of Cr-doped ${\rm V_{2}O_{3}}$
for the insulating phase of pure ${\rm V_{2}O_{3}}$ is justified
by the observation that Cr-doping is equivalent to the application
of (negative) pressure.

\bibitem{mattheiss94} L.\ F.\ Mattheiss, J.\ Phys.: Cond.\ Matt.\ {\bf 6},
6477 (1994).

\bibitem{Solovyev} I. Solovyev, N. Hamada, K. Terakura,
Phys.\ Rev.\ B {\bf 53}, 7158 (1996).

\bibitem{Anisimov91}  V. I. Anisimov, J. Zaanen, and O. K. Andersen, Phys.
Rev. B {\bf 44}, 943 (1991); V. I. Anisimov, F. Aryasetiawan, and
A. I. Lichtenstein, J. Phys. Cond. Matter {\bf 9}, 767 (1997).

\bibitem{Foot3} All QMC results were obtained for $T=1000$K due to the
limitations in available the computing resources\cite{Nekrasov00}.

\bibitem{U-exp} A. T. Mizokawa, A. Fujimori, Phys.Rev.B {\bf 48}, 14150
(1993); J. Zaanen, G. A. Sawatzky, J. Solid State Chem., {\bf 88},
8 (1990).

\bibitem{Schramme00} M. Schramme, Ph.D. thesis, Universit\"{a}t Augsburg,
2000;
 M.~Schramme {\it
et al.}, in preparation.

\bibitem{Mueller} O. M\"{u}ller, J. P. Urbach, E. Goering, T. Weber,
R. Barth, H. Schuler, M. Klemm, M. L. denBoer, and S. Horn,
Phys. Rev. B {\bf 56}, 15056
(1997).

\bibitem{S-exp} D.J. Arnold, R.W. Mires, J. Chem. Phys. {\bf 48}, 2231 (1968)

\bibitem{brown} P.J. Brown, M.M.R. Costa, K.R.A. Ziebeck, J.Phys.: Condens.
Matter {\bf 10}, 9581 (1998).

\bibitem{Majumdar94} P. Majumdar and H.R. Krishnamurthy, Phys. Rev. Lett.
{\bf 73}, 1525 (1994).

\bibitem{KVC} J.W. Allen and R.M. Martin, Phys. Rev. Lett. {\bf 49}, 1106,
(1982); M. Lavagna, C. Lacroix, and M. Cyrot, J. Phys. F {\bf 13},
1007 (1983);  C.~Huscroft, A. K. McMahan, and R. T. Scalettar,
 Phys.~Rev.~Lett.~{\bf 82},
2342 (1999); K. Held, C. Huscroft, R.T. Scalettar, A. K. McMahan,
 Phys. Rev. Lett. {\bf 85}, 373
(2000).
\end{references}
\end{document}